\documentclass[notitlepage,aps,prl,reprint,twocolumn,longbibliography,superscriptaddress]{revtex4-1}
\usepackage{graphicx}
\usepackage{amsmath}
\usepackage{amssymb}
\usepackage{comment}
\usepackage[colorlinks, allcolors=blue]{hyperref}
\usepackage[all]{hypcap}
\usepackage[mathlines]{lineno}
\usepackage{braket}
\usepackage{wrapfig}
\usepackage{lipsum}
\usepackage{ulem}

\newcommand{\pref}[2]{\hyperref[#1]{\ref{#1}(#2)}}
\newcommand{\preff}[2]{\hyperref[#1]{\ref{#1}#2}}
\newcommand{\eqpref}[1]{\hyperref[#1]{(\ref{#1})}}

\newcommand{\squig}{{\raise.17ex\hbox{$\scriptstyle\sim$}}}

\begin{document}
\title{Nonlinear dynamics in a synthetic momentum state lattice}
\author{Fangzhao Alex An}
\thanks{Present address: Honeywell Quantum Solutions, Golden Valley, MN 55422}
\affiliation{Department of Physics, University of Illinois at Urbana-Champaign, Urbana, IL 61801-3080, USA}
\author{Bhuvanesh Sundar}
\affiliation{Department of Physics and Astronomy, Rice University, Houston, TX 77005, USA}
\affiliation{Rice Center for Quantum Materials, Rice University, Houston, TX 77005, USA}
\affiliation{JILA, Department of Physics, University of Colorado, Boulder, CO 80309, USA}
\author{Junpeng Hou}
\affiliation{Department of Physics, The University of Texas at Dallas, Richardson, Texas 75080-3021, USA}
\author{Xi-Wang Luo}
\affiliation{Department of Physics, The University of Texas at Dallas, Richardson, Texas 75080-3021, USA}
\author{Eric J.~Meier}
\affiliation{Department of Physics, University of Illinois at Urbana-Champaign, Urbana, IL 61801-3080, USA}
\author{Chuanwei Zhang}
\email{chuanwei.zhang@utdallas.edu}
\affiliation{Department of Physics, The University of Texas at Dallas, Richardson, Texas 75080-3021, USA}
\author{Kaden R.~A.~Hazzard}
\email{kaden@rice.edu}
\affiliation{Department of Physics and Astronomy, Rice University, Houston, TX 77005, USA}
\affiliation{Rice Center for Quantum Materials, Rice University, Houston, TX 77005, USA}
\author{Bryce Gadway}
\email{bgadway@illinois.edu}
\affiliation{Department of Physics, University of Illinois at Urbana-Champaign, Urbana, IL 61801-3080, USA}
\date{\today}

\begin{abstract}
The scope of analog simulation in atomic, molecular, and optical systems has expanded greatly over the past decades.
Recently, the idea of synthetic dimensions -- in which transport occurs in a space spanned by internal or motional states coupled by field-driven transitions -- has played a key role in this expansion.
While approaches based on synthetic dimensions have led to rapid advances in single-particle Hamiltonian engineering, strong interaction effects have been conspicuously absent from most synthetic dimensions platforms.
Here, in a lattice of coupled atomic momentum states, we show that
atomic interactions
result in large and qualitative changes to dynamics in the synthetic dimension.
We explore how the interplay of nonlinear interactions and coherent tunneling enriches the dynamics of a one-band tight-binding model, giving rise to macroscopic self-trapping and phase-driven Josephson dynamics with a nonsinusoidal current-phase relationship, which can be viewed as stemming from a nonlinear band structure arising from interactions.
\end{abstract}
\maketitle

The concept of synthetic dimensions~\cite{Review-OzawaPrice}, where motion in space is abstracted to encompass dynamics in spaces spanned by internal~\cite{Boada-SynthDim,Celi-SynthSynth,Wall-Synthetic-Clock,Sundar-Synth-Mol,Sundar2} or discrete motional states~\cite{Gadway-KSPACE,Price-Shaking,Lincoln1}, has led to new capabilities for analog simulation with quantum matter~\cite{Mancini2015,Stuhl2015,Livi2016,Kolkowitz2017,Shin-Tubes1,Shin-Tube2,Chen-Tubes1,Meier-AtomOptics,Meier-SSH,Meier-TAI,An-FluxLadder,An-Disorder,An-MobEdge,An-GAA,Lapp_2019,Xiao2020,Gou-NonRecip,Xie-TopQuantWalks,Genkina_2019,Sugawa1429,chen2020quantum,Xiao-Top-Critical,Chalopin2020,RydSynth,Cornish-synth}.
Under this approach, the role of synthetic lattice sites is played by discrete internal or motional states, and tunneling between the sites is accomplished by driving state-to-state transitions. The spectroscopic control over the resulting tight-binding model, including direct control over tunneling phases, opens up new capabilities for Hamiltonian engineering. By combining these new possibilities for single-particle control with the native interactions in atomic and molecular systems~\cite{Sundar-Synth-Mol,Sundar2,An-Inter,Hou-PRL,Barbarino_2016,Guan-MSL,GuanSwallow}, synthetic dimensions hold much promise for the exploration of exotic many-body physics.

To date, studies based on synthetic dimensions have almost uniformly been restricted to the non-interacting regime, or regimes in which there are mostly subtle modifications to the system's behavior~\cite{Bromley-Inter,An-MobEdge,An-GAA,Xie-TopQuantWalks}. Here, using lattices of atomic momentum states, we observe that the dynamics in the synthetic dimension is qualitatively altered by the presence of nonlinear atomic interactions.
The controlled addition of atomic interactions to even a simple one-band tight-binding model leads to a rich variety of phenomena, including the transition to a macroscopically self-trapped regime and the appearance of a nonsinusoidal current-phase relationship in phase-driven Josephson dynamics, which we observe with single-site resolution in the synthetic dimension.
We note that these observations can be viewed as resulting from the emergence of a nonlinear band structure arising from interactions~\cite{Erich-swallowtails,Koller-swallowtails}.
As synthetic lattices can readily incorporate designer disorder, gauge fields, and other features, these results pave the way for future explorations of novel collective phenomena in synthetic dimensions.

\begin{figure}[t]
	\includegraphics[width=0.97\columnwidth]{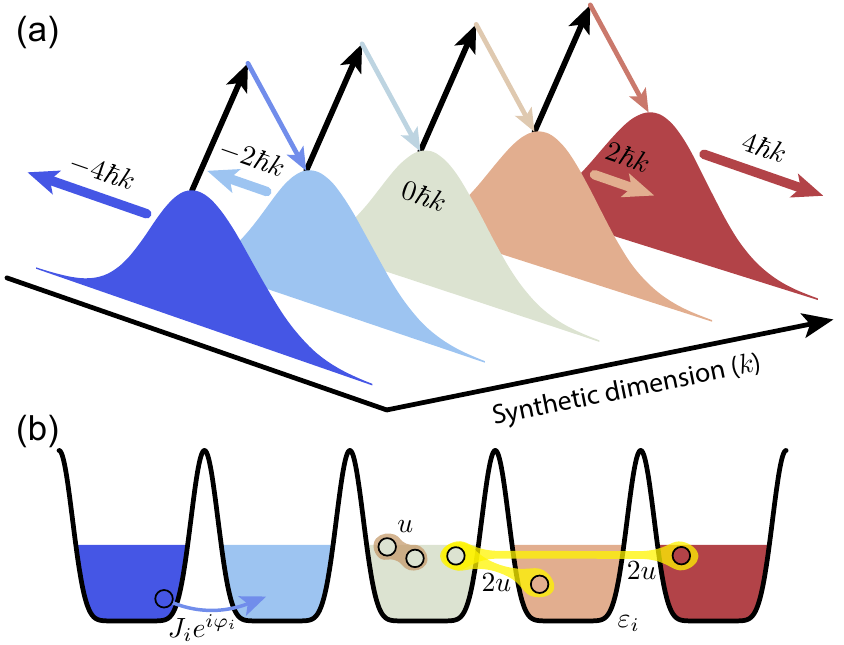}
	\caption{\label{FIG:fig1}
		\textbf{Synthetic momentum state lattices with atom-atom interactions.}
		\textbf{(a)}~Two-photon Bragg transitions individually couple atomic momentum modes along a synthetic dimension.
		\textbf{(b)}~The synthetic lattice features inter-site tunneling terms with controlled amplitude $J_i$ and phase $\varphi_i$
		set by the Bragg laser amplitudes and phases, lattice site energies $\varepsilon_i$ set by the two-photon Bragg detuning,
		and naturally occurring long-ranged and mode-dependent atomic interactions: atoms in the same site experience a repulsive pairwise interaction of strength $u$, and atoms in different sites have a repulsive interaction of strength $2u$.
	}
\end{figure}

Figure~\ref{FIG:fig1} depicts our approach based on atomic momentum states~\cite{Gadway-KSPACE,Meier-AtomOptics}.
A Bose--Einstein condensate (BEC) containing $\squig 10^5$ $^{87}$Rb atoms is subjected to two counterpropagating lasers (wavelength $\lambda = 1064$~nm, wave vector $k = 2\pi/\lambda$) that both trap the atoms and couple the atomic momentum states.
These laser fields allow for discrete momentum transfer to the atoms in units of two photon recoil momenta, $2\hbar k$, defining a set of states with momenta $p_n = 2 n \hbar k$ and energies $E_n = p_n^2 / 2 M$, with $M$ the atomic mass.
Because the energy difference between adjacent states $p_n$ and $p_{n+1}$ is unique, we individually address various nearest-neighbor transitions simply by driving the atoms with a comb of discrete frequency components.
We write a controlled spectrum of frequency components onto one of the lasers, such that the interference between the multifrequency beam (colored arrows) and the single-frequency beam (black arrows) addresses two-photon Bragg transitions between many pairs of momentum states.
This creates a lattice of states in a synthetic dimension (momentum), where all tunneling amplitudes, tunneling phases, and site energies are controlled by the amplitudes, phases, and detunings from Bragg resonance of the corresponding laser frequency components~\cite{Meier-AtomOptics}.
The site populations are all simultaneously measured via time-of-flight absorption imaging.

We briefly summarize how atomic collisions lead to relevant interactions in synthetic lattices of atomic momentum states~\cite{An-Inter,Ozeri-RMP}. The strength of the collisional interaction is $U = g \rho = \left( 4\pi \hbar^2 a / M \right) \rho$,
where $a$ is the $s$-wave scattering length ($\sim$100~$a_0$ for $^{87}$Rb), and $\rho$ is the real-space atomic density.
In momentum-space, the $s$-wave collision is ostensibly all-to-all and state-independent. However, quantum statistics leads to a natural state-dependence to the interactions.
A pair of atoms occupying the same momentum state experience a direct interaction term $u = U/N$. In contrast, a pair of atoms occupying distinguishable momentum states experience an additional exchange energy due to bosonic exchange statistics, resulting in a total interaction of $2u$ [Fig.~\pref{FIG:fig1}{b}].

\begin{figure}[hbt!]
	\includegraphics[width=0.97\columnwidth]{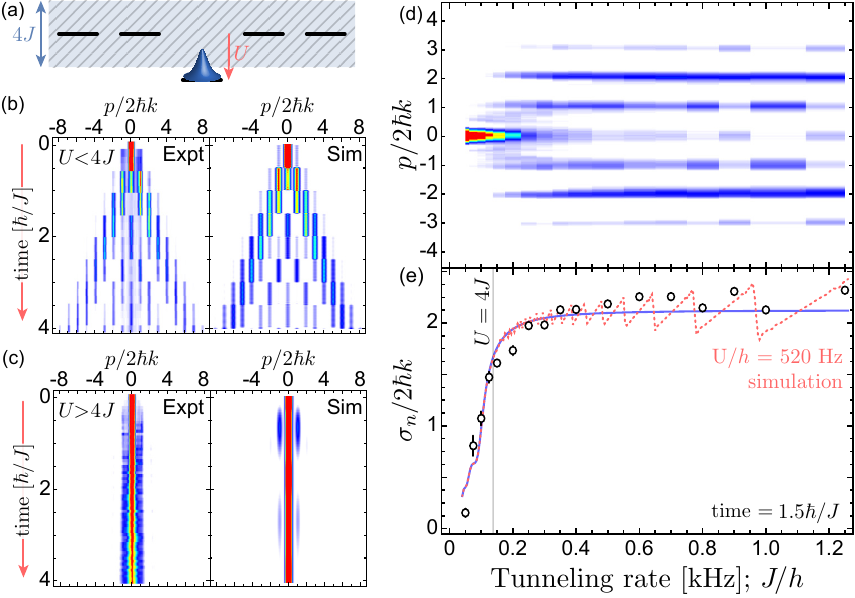}
	\centering
	\caption{\label{FIG:fig2}
		\textbf{Macroscopic self-trapping in an array of laser-coupled momentum states.}
        \textbf{(a)}~Atoms initialized to a single lattice site evolve in a one-dimensional lattice with uniform tunneling $J$. Self-trapping occurs when the collective interaction strength $U$ exceeds the tunneling bandwidth $4J$.
		\textbf{(b,c)}~Population dynamics for strong ($J/h = 1281(5)$~Hz) and weak tunneling ($J/h = 93.3(2)$~Hz). Left: Data from integrated optical density (OD) images after $18$~ms time of flight, averaged over 5 experimental realizations. Right: Dynamics from numerical simulations.
		\textbf{(d)}~Integrated OD images plotted vs.\ the tunneling strength $J$, taken at times of $t = 1.5 \hbar / J$.
		\textbf{(e)}~Standard deviation of the atomic distributions from (d), shown alongside the simulation results. The solid blue curve assumes the ideal Hamiltonian after application of the rotating wave approximation, while the dashed red curve includes residual time dependence due to off-resonant effects. The expected self-trapping transition $U/J=4$ is shown as a vertical gray line.
		Data for (d,e) are averaged over 20 experimental realizations, and the error bars in (e) are the standard error of the mean.
		Numerical simulations in (b,c,e) assume a homogeneous mean-field energy of $U/h = 520$~Hz.
	}
\end{figure}

In Ref.~\cite{An-Inter}, we laid the groundwork for interactions in momentum-state lattices and their influence on a double-well system.
Here, we explore the simplest one-dimensional lattice, with uniform nearest-neighbor tunneling, $J$, between 21 sites.
Following the above discussion, the interaction is approximately~\cite{SuppMats} described by $ u \sum_i n_i (n_i - 1)/2 + 2u \sum_{i<j} n_i n_j = U(N-1/2) - (u/2) \sum_i n_i^2$. This yields, up to irrelevant energy shifts, the combined Hamiltonian
\begin{equation}
H = -J\sum_i ( c_i^\dagger c^{\phantom \dagger}_{i+1} + h.c.) - \frac{u}{2}\sum_i n_i^2 \ .
\end{equation}
Here, $i$ and $j$ index the synthetic lattice sites and $n_i = c_i^\dag c^{\phantom \dagger}_i$ is the number operator for site $i$, with $c^{\phantom \dagger}_i$ and $c^\dag_i$ the site-$i$ annihilation and creation operators, respectively.

We begin with all atoms in the nearly pure BEC and compose the Bragg frequencies such that the
$k=0$ mode relates to the central lattice site, as depicted in Fig.~\pref{FIG:fig2}{a}.
We note that, because each synthetic site hosts only one discrete state, the lattice is intrinsically single-band, with a simple cosine energy dispersion and a total bandwidth of $4J$.
Initializing at a single site projects the atoms onto a superposition of all eigenstates in the band structure, giving rise to nonequilibrium dynamics and ballistic transport across the lattice~\cite{Meier-AtomOptics,An-Disorder}, as shown in Fig.~\pref{FIG:fig2}{b} in the limit where tunneling ($J/h = 1281(5)$~Hz) exceeds the collective interaction shift ($U/h = 520$~Hz).

We expect that, at weaker $J$, the nonlinear interactions can compete with the tunneling and influence the dynamics. In particular, when the nonlinear interaction strength $U$ becomes greater than the full tunneling bandwidth $4J$, the atoms should be confined from spreading in a soliton-like state. That is, tunneling from the initially occupied mode will be suppressed by being off-resonant in energy with the other modes, a mechanism known as macroscopic quantum self trapping~\cite{Raghavan,Maciej-self-trap}. Figure~\pref{FIG:fig2}{c} shows such a lack of dynamics in the weak-tunneling regime ($J/h = 93.3(2)$~Hz), with atoms largely remaining on the initially populated site.
Dynamics in both the strong- and weak-tunneling limits are in good agreement with Gross--Pitaevskii equation (GPE) simulations, which, for simplicity, approximate the trapped condensate as having a uniform mean-field energy of $U/h=520$~Hz~\cite{SuppMats}.
Figure~\pref{FIG:fig2}{d} shows the transition from ballistic spreading to self-trapping more comprehensively for a range of tunneling rates. All data are taken at equivalent evolution times of $t = 1.5 \ \hbar/J$.
For large $J$, the ballistic evolution leads to a bimodal distribution peaked around the $\pm 4 \hbar k$ states.
As the tunneling rate decreases and interactions begin to dominate, population is found to spread less far out, eventually remaining fully within the initialized site in the self-trapping region.
By plotting the distribution width $\sigma_n$ for each $J$ in Fig.~\pref{FIG:fig2}{e}, we see that it is nearly constant at large $J$ and that there is a rather steep turnover at low $J$ values, in good agreement with the GPE simulations.

\begin{figure*}[htb!]
	\includegraphics{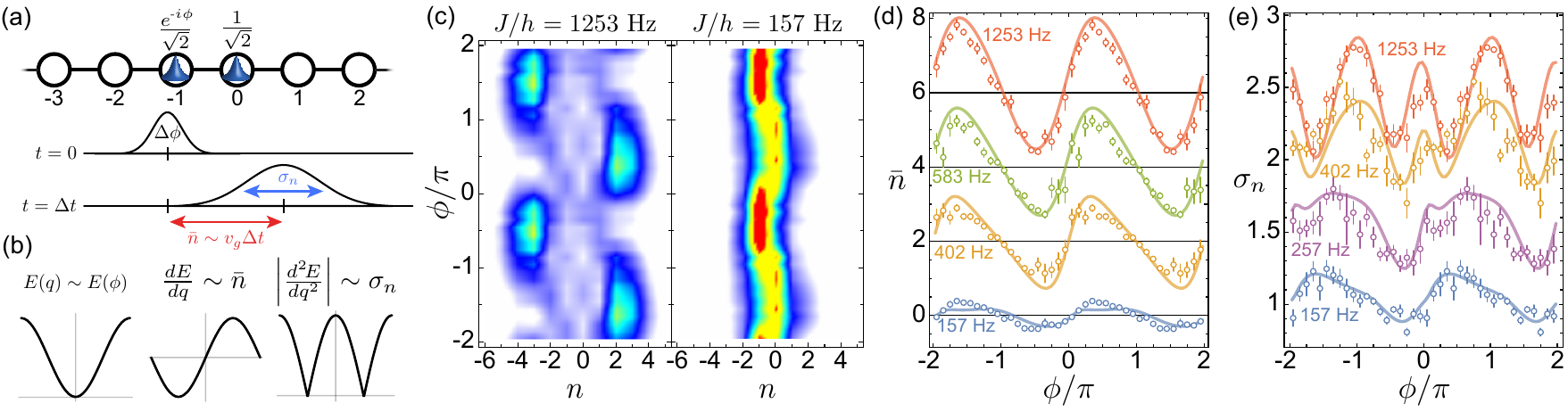}
	\centering
	\caption{\label{FIG:fig4}
		\textbf{Phase-driven dynamics in a synthetic bosonic Josephson junction array.}
        \textbf{(a)}~Atoms are prepared in a superposition state on two lattice sites, $n = -1$ and $n = 0$, with equal weight and a relative phase of $\phi$. Starting at time $t=0$, the full lattice is turned on and the atoms evolve for a time $\Delta t$. We then measure the atomic distribution, from which we determine the mean displacement $\bar{n}$ (mean position + 0.5) and spread  $\sigma_n$ (standard deviation of position) of the wavepacket.
        \textbf{(b)}~In the non-interacting limit, $\bar{n}$ and $\sigma_n$ can be related to the single-band energy $E(q)$ of quasimomentum states in the synthetic lattice, as the wavepackets with phase offset $\phi$ populate a spread of quasimomentum states about a mean synthetic quasimomentum $q_0 = \phi$.
        The plots show the band energy $E(q)$, the group velocity $v_g \propto dE/dq$, and the dispersion $d^2 E / dq^2$ of the synthetic lattice.
        \textbf{(c)}~Integrated density patterns (after 2 tunneling times) vs.\ $\phi$ for the regimes of strong (left, $J/h = 1253(12)$~Hz) and weak tunneling (right, $J/h = 156.7(1.3)$~Hz).
		\textbf{(d,e)}~$\bar{n}$ vs.\ $\phi$ and $\sigma_n$ vs.\ $\phi$, respectively, for several tunneling values [$J/h = \{156.7(1.3), 257(3), 402(3), 583(6), 1253(12)\}$]. Dots are experimental data and the lines are theory.
		The results for different tunneling values are offset for clarity in (d). No offset is applied for (e).
		The theory curves in (d,e) are based on a real-space GPE simulation, as described in the text.
		Data for (c-e) are averaged over 10 experimental realizations, with error bars in (d,e) representing the standard error of the mean.
		The data sets in (c-e) were taken over a $2\pi$ range of $\phi$ values, uniformly shifted in phase to correct for phase shifts acquired during the wavepacket initialization, and then mapped onto a larger and redundant $4\pi$ range.
	}
\end{figure*}

This direct, site-resolved observation of macroscopic self trapping, in excellent agreement with the Gross-Pitaevskii mean-field simulations~\cite{SuppMats}, underscores the coherent interplay between tunneling dynamics and atomic interactions in our synthetic lattice. We note that, while there have been numerous explorations of self-trapping~\cite{Albiez-DirectJosephson,Zibold,Chen-DoubleWell,Fattori-TunableDoubleWell,Anker-SelfTrap,Reinhard-SelfTrap,OTT-ndc,Xhani_2020,Smaleeaax1568} and other Josephson phenomena~\cite{Anderson-Interference,Cataliotti-01,Levy-JosephsonBEC,LeblancJunction,Jorg-Relaxation,Roati-JosephsonBECBCS,2D-JosephsonArray,Steinberg-tunneling} in real-space potentials and in few-state spin mixtures~\cite{Chapman-Spinor,Zibold,Ferrari-Internal,Gerbier-Shapiro}, this is the first such observation in a many-site synthetic lattice. The incredible control over synthetic lattices promises the extension to future explorations of more complex dynamical phase transitions~\cite{Heyl_2019}, including those of non-ground states~\cite{Tian-excitedstate}.

We now move to make use of one of the most immediate and unique tools of synthetic lattices, the direct control of tunneling phase~\cite{Hou-PRL}. Specifically, we utilize this phase control to initialize superposition states with atoms delocalized over two adjacent sites and with a relative phase $\phi$, as illustrated in Fig.~\pref{FIG:fig4}{a}. We then probe the phase-driven Josephson dynamics~\cite{JOSEPHSON1962251,Raghavan} in our many-site lattice by suddenly turning on the tunnelings and allowing the initialized wave packet to evolve for two tunneling times.
As in the case of single-site initialization, there will be a propensity for the atoms to spread out across the lattice due to coherent tunneling. In addition, the relative phase $\phi$ can act to drive a net current of atoms in the lattice. In this case of phase-driven dynamics, we again expect interactions to play a nontrivial role.

We can gain some insight into the expected phase-dependent dynamics of our initial two-site wave packets by considering their far-field response, and by considering their distribution after two tunneling times as an approximation thereof.
We expect the average displacement $\bar{n}$ in the lattice after the quench time to be roughly proportional to the group velocity $v_g$ of the initialized wave packet. Correspondingly, the spread (standard deviation) of the distribution after the quench, $\sigma_n$, should correspond to the spread in group velocities of the initialized wave packet. In the absence of interactions, the group velocity and dispersion (spread in group velocity) of our initialized two-site wave packets would be determined by the simple cosine band structure of the synthetic tight-binding lattice. That is, the initialization procedure will populate a range of Bloch-like states with an average synthetic quasimomentum of $q_0 = \phi$.

As shown in Fig.~\pref{FIG:fig4}{a}, the measured displacement and spread of the final distribution thus serve as probes of the synthetic band structure, at least in the non-interacting limit. Figure~\pref{FIG:fig4}{b} provides an intuitive picture for the expected phase-driven dynamics in the $U=0$ limit.
The measured displacement $\bar{n}$, proportional to the group velocity $v_g$, should scale as the first derivative of the band structure, \textit{i.e.}, with $\partial E / \partial q \propto \sin (\phi)$. More generally (away from $U=0$), measurement of $\bar{n}$ vs.\ $\phi$ probes the current-phase relationship (CPR)~\cite{CPR-RMP} of this synthetic bosonic Josephson junction array.
Similarly, in the $U=0$ and far-field limit, one expects the measured width $\sigma_n$ to scale as the magnitude of the second derivative of the band structure, \textit{i.e.}, with $|\partial^2 E / \partial q^2| \propto |\cos (\phi)|$.

Figure~\pref{FIG:fig4}{c} depicts the evolved atomic distributions in the tunneling-dominated regime (left, $J/h = 1253(12)$~Hz) and interaction-dominated regime (right, $J/h = 156.7(1.3)$~Hz), with plots of the integrated density patterns vs.~$\phi$. In the large $J$ limit, where interactions play a minor role, we indeed see agreement with the above description: Atomic currents are driven in the positive (negative) direction when the sinusoidal group velocity is maximally positive (negative) for $\phi$ values of $\pi/2$ ($-\pi/2$). At $\phi$ values of $0$ and $\pi$, we find atoms flowing equally in the positive and negative direction, and the distribution has maximum width. In lowering $J$ and entering into the interaction-dominated regime (right panel), we find dramatically different atomic distributions, with an almost complete collapse and absence of spreading, similar to the earlier self-trapped condition. However, we find that a clear $\phi$-dependence of the displacement survives, hinting at the formation of a partially mobile soliton.

A more quantitative analysis of the phase-driven dynamics is found in Fig.~\pref{FIG:fig4}{d,\,e}, with the $\phi$-dependence of $\bar{n}$ and $\sigma_n$
shown for several $J$ values.
Even for the largest tunneling
($J/h = 1253(12)$~Hz), we observe a clear deviation from the $U=0$ expectation.
The measured CPR ($\bar{n}$ vs.\ $\phi$) deviates from the simple sine dependence, instead showing a skewed form that is in good agreement with the numerical simulation curves, which are based upon real-space GPE simulations (solid line)~\cite{SuppMats}.
Similarly, while in the non-interacting limit
$\sigma_n$ is $\pi$-periodic with $\phi$ due to the insensitivity to the sign of the dispersion, we see that the distribution width is larger at the band edge ($\phi = \pm\pi$) than at the band center ($\phi = 0$).

While the density and scattering length of our atomic gas are fixed, we can increase the ratio $U/J$,
to explore how the phase-driven dynamics are impacted as interactions play a more dominant role, by simply lowering
the tunneling $J$ (maintaining an evolution time of $2\hbar/J$).
As $J$ is decreased ($U/J$ increased), we observe that the measured current-phase response starts to become further skewed and nonsinusoidal.

For a general tunnel junction, a nonsinusoidal CPR can arise for a host of reasons~\cite{CPR-RMP}.
In real materials~\cite{CPR-RMP,Hart-Interactions,Klapw-PiJunction,Kouw-PiJunction,PiJunction-nanotube} and in real-space atomtronic junctions~\cite{Ramanthan-WeakLink,Campbell-Hysteresis,EckelCPR,Ryu-SQUID,Chien2015,Neely-Atomtronic,Ryu2020}, such a skewed CPR can arise even in the absence of interactions between the current carriers, purely from the details of the junction.
In our synthetic junction array,
however, the ``junctions'' are featureless, simply relating to laser-driven state-to-state transitions.
Still, even in this ideal synthetic junction array, it is predicted that nonlinear interactions alone can give rise to a skewed, nonsinusoidal CPR, as seen from the theory comparisons for increasing $U/J$~\cite{SuppMats}.
In addition to becoming skewed, the overall amplitude of the CPR becomes suppressed as $U/J$ increases. This muted response
signals that the atoms of the fluid become effectively ``heavier'' as the collective interactions are increased relative to the tunneling.
This can be interpreted in terms of a collapse of the effective nonlinear band structure
as the atoms' effective mass increases and they behave like a collective soliton.

The dependence of the distribution widths $\sigma_n$ on $J$ can further shed light on how the atomic mobility is impacted by the increase in $U/J$.
We observe that the overall ($\phi$-averaged) values of the $\sigma_n$ slowly decrease for small $U/J$ values (shown for the cases of $J/h = 402(3)$~Hz and 1253(12)~Hz). Then, for larger $U/J$ values, there is a sudden collapse of the distribution widths, similar to that seen in Fig.~\pref{FIG:fig2}{e}.
The dependence of $\sigma_n$ on $\phi$ reveals additional features. The initial asymmetry in the $\sigma_n$ values at $\phi = 0$ and $\pm\pi$ becomes more pronounced as the tunneling is decreased. Specifically, we find that the sudden reduction in $\sigma_n$ with decreasing $J$ (increasing $U/J$) first occurs for the $\phi=0$ peak feature, and then, at lower $J$ and larger $U/J$, for the broader $\phi=\pm\pi$ peaks.
The GPE simulations show a similar response, with the central $\sigma_n$ peak getting narrower and smaller before the decrease at $\phi = \pm \pi$. This signals a $\phi$-dependent collapse of transport, with the atoms first becoming self-trapped for equal-phase superpositions ($\phi=0$) and then, at larger $U/J$, for the out-of-phase state ($\phi=\pi$). This behavior in our many-site array is reminiscent of the frequency response seen in Josephson double-wells, where in-phase superposition modes are softened by the introduction of nonlinear interactions, while the out-of-phase plasma mode acquires a stiffened response~\cite{Raghavan,Fattori-TunableDoubleWell}.

We note that the numerical simulation curves in Fig.~\pref{FIG:fig4}{d,\,e} are based on real-space GPE calculations~\cite{SuppMats}, which naturally incorporate the influence of the real-space trapping potential and the resulting inhomogeneous atomic density. This real-space treatment was necessitated by the use of a stiffer trapping potential, characterized by a trapping frequency of 60~Hz along the spatial axis of the synthetic lattice (as compared to $\sim$10$-$15~Hz for the data of Fig.~\ref{FIG:fig2}). This stiffer trap additionally led to a larger characteristic mean interaction strength, with a value of $\bar{U}/h = 1568$~Hz used in the simulation curves of Fig.~\pref{FIG:fig4}{d,\,e}. Importantly, the key observations of a non-sinusoidal CPR and a $\phi$-dependent collapse of the distribution width are both found in simulations that assume a homogeneous $U$ term~\cite{SuppMats}. However, the real-space GPE analysis is key to capturing several features, or lack thereof, in the experimental data. Specifically, the effective spatial averaging over a range of $U/J$ leads to an overall dulling of the $\phi$-dependent collapse of $\sigma_n$. Similarly, GPE simulations with a uniform $U$ predict an intriguing reversal of the atomic current (sign reversal of the CPR) over a range of $U/J$ values close to the onset of self-trapping, however this effect is not present in the real-space simulations. Finally, the slow drift of the data minima away from $\phi = 0$ in Fig.~\pref{FIG:fig4}{e}, seen for increasing $U/J$, stems from the real-space trapping potential and our exploring dynamics out to times that are not small compared to the trap period.

To conclude, we have performed the first measurements of significant interaction effects in a many-site synthetic lattice.
These observations pave the way for the exploration of exotic nonlinear phenomena by straightforward extensions to more sophisticated synthetic lattices with gauge fields, disorder, and non-Hermiticity.

In addition to nonlinear phenomena captured by a classical mean-field description, atomic collisions~\cite{Ozeri-RMP} can also give rise to quantum correlations in synthetic momentum state lattices, which can be explored through number-resolved imaging~\cite{Luo620,Gerb}.
Furthermore, the extension of synthetic dimensions studies to spin systems~\cite{Sundar-Synth-Mol,Sundar2,RydSynth} opens up a promising path to studying strongly correlated matter in synthetic dimensions.

\section{Acknowledgements}
This material is based upon work supported by the Air Force Office of Scientific Research under Grant No.~FA9550-18-1-0082. K.R.A.H and B.S. acknowledge funding support from NSF Grant No.~PHY-1848304 and Welch Grant No.~C-1872. J.H., X.W.L. and C.Z. acknowledge funding support from AFOSR (FA9550-20-1-0220), NSF (PHY-1806227), and ARO (W911NF-17-1-0128).

\bibliographystyle{apsrev4-1}

\newcommand{\hH}{\hat{H}}
\newcommand{\hp}{\hat{\psi}}
\newcommand{\+}{^\dagger}
\newcommand{\rmi}{{\rm i}}
\newcommand{\rmd}{{\rm d}}
\newcommand{\br}{\mathbf{r}}
\newcommand{\vk}{\mathbf{k}}
\newcommand{\vp}{\mathbf{p}}
\newcommand{\vq}{\mathbf{q}}

\clearpage

\section*{Supplementary Materials}
\renewcommand{\theequation}{S\arabic{equation}}
\renewcommand{\thefigure}{S\arabic{figure}}
\setcounter{equation}{0}
\setcounter{figure}{0}

\subsection{Experimental details}

These experiments based on condensates of $^{87}$Rb operate with a fixed atomic scattering length $a$ and a nearly fixed particle number. For the studies in the main text (Fig.~\ref{FIG:fig2} and Fig.~\ref{FIG:fig4}), the characteristic collective interaction strength $U$ is only weakly altered by a change of the trapping potential, whereby the density is increased for the data of Fig.~\ref{FIG:fig4} (with a stiffer trap having a trapping frequency of 60~Hz along the direction of imparted momentum) as compared to Fig.~\ref{FIG:fig2} (based on a looser, essentially single-beam trap having a trapping frequency of $\sim$10$-$15~Hz along the direction of imparted momentum). A larger variation of the ratio of Hamiltonian parameters $U$ and $J$ is achieved by directly modifying the tunneling energy $J$ and investigating dynamics at equivalent times in units of the tunneling time $\hbar/J$.

\subsubsection{Tunneling calibration}

For all experimental data runs, we calibrated the value of tunneling by independently measuring two-site Rabi oscillations.
That is, we turn on a single tunneling link for a range of evolution times and fit the resulting Rabi oscillation dynamics to obtain the effective tunneling rate. This is performed in the strong-tunneling limit, to minimize the influence of the mean-field interactions on the observed rate of tunneling.
We then generate the frequencies for a lattice experiment under identical experimental conditions (voltages, laser powers), scaling the tunneling amplitudes according to the calibration.
We run calibrations immediately before and after each data set, and the quoted tunneling values relate to the mean of the two calibration values, to account for drifts over time. The quoted tunneling errors relate to the standard error of the mean, incorporating the uncertainty (standard error of the fit of the Rabi oscillations) for each individual calibration.

\subsubsection{Initialization of two-site superpositions}

For the data in Fig.~\ref{FIG:fig4}, we initialized the system with equal populations at sites $0$ and $-1$.
We apply a step-function $\pi/2$ pulse to the $0 \leftrightarrow -1$ tunneling link (evolving for one quarter of a Rabi period), also applying a controlled phase shift within the range $\phi/\pi = [-1,0.9]$ (in steps of 0.1).
In all cases, the two-site superposition states acquire an additional relative phase shift of $-0.85\pi$ due to interactions during the initialization process, as determined by a comparison of the results to simulations in the high-tunneling limit.
This leaves us with initialized states characterized by relative phase terms $\phi/\pi = [-0.95,0.95]$.
For presentation purposes, we map the data onto a redundant range with span $4\pi$, copying data in the range $\phi/\pi = [0.05,0.95]$ to $\phi/\pi = [-1.95,-1.05]$ and copying data in the range $\phi/\pi = [-0.95,-0.05]$ to $\phi/\pi = [1.05,1.95]$.

\begin{figure*}[t!]
	\includegraphics[width=0.75\textwidth]{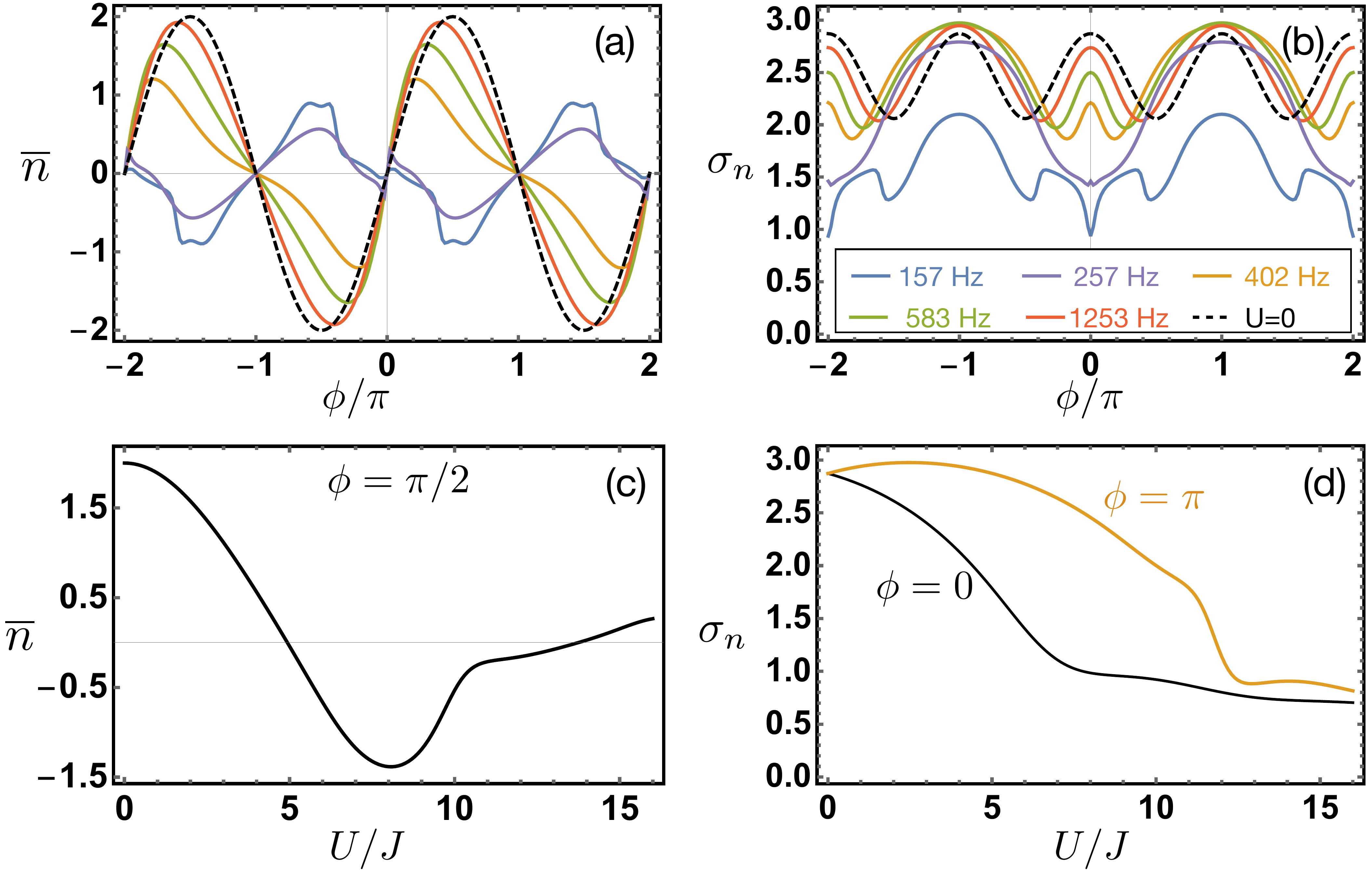}
	\centering
	\caption{\label{FIG:figS1}
		\textbf{Discrete-mode Gross--Pitaevskii simulations of phase-driven dynamics in a 21-site lattice.}
		\textbf{(a,b)}~Mean displacement $\bar{n}$ and distribution width $\sigma_n$ vs. the initial relative phase $\phi$ of two-site superposition states after an evolution for two tunneling times. The curves relate to the same quantities presented in Fig.~\ref{FIG:fig4}, but are based on simplified discrete-mode GP simulations that neglect the influence of the real-space trapping potential. The solid curves assume a homogeneous mean-field energy of $U/h = 1500$~Hz, and are evaluated for the tunneling values labeled in the legend in (b). For reference, dashed black lines denote the mean displacement $\bar{n}$ and distribution width $\sigma_n$ vs. $\phi$ results for the non-interacting model.
		\textbf{(c)}~The mean displacement $\bar{n}$ of the superposition state with an initial relative phase of $\pi/2$, plotted as a function of the interaction energy-to-tunneling energy ratio $U/J$ (for an evolution over two tunneling times).
		\textbf{(d)}~The distribution width $\sigma_n$ of the superposition state with initial relative phases of $0$ (black) and $\pi$ (orange), plotted as a function of the interaction energy-to-tunneling energy ratio $U/J$ (for an evolution over two tunneling times).
	}
\end{figure*}

\subsection{Numerical simulations}

We compare the experimental data to simulations based on the nonlinear Schr\"{o}dinger equation (NLSE), alternatively referred to in the text as the Gross-Pitaevskii equation (GPE), with mean-field nonlinearities capturing the influences of atomic interactions. We utilize simulations of varying levels of complexity, opting for the simplest description capable of capturing the key experimental observations for each study. These analyses involve (1) solving a simple NLSE of 21 discrete modes with a uniform interaction energy to capture the self-trapping transition observed in Fig.~\ref{FIG:fig2} and (2) solving a real-space 1D NLSE that additionally incorporates the influences of the real-space trapping potential and the inhomogeneous and time-evolving atomic density to capture the phase-driven dynamics of Fig.~\ref{FIG:fig4}.

\subsubsection{Discrete-mode Gross-Pitaevskii simulations}

The simulation curves accompanying the experimental data in Fig.~\ref{FIG:fig2} reflect the results of a NLSE with 21 discrete momentum modes. This classical mean-field simulation is justified by the macroscopic populations of the momentum modes in the experiments, with the number of atoms greatly exceeding the number of coupled momentum modes. For concreteness, we solve the dynamics of the (normalized) wave functions under the equations
\begin{equation}
i \hbar \dot{\psi_j} = - J (\psi_{j-1} + \psi_{j+1}) + U \sum_{j'} g_{j,j'} |\psi_{j'}|^2 \psi_{j} \ ,
\label{EQ:PsiHam}
\end{equation}
where $g_{j,j'} = 2 - \delta (j - j')$. This term reflects the different strengths of intra-mode and inter-mode interactions~\cite{Ozeri-RMP}. The exact difference by a factor of 2 reflects a simplifying approximation, whereby we assume the momentum modes to be perfectly distinguishable. This approximation, which neglects the influence of screening, is strictly valid when $2U \ll 4E_r \approx h \times 8.10$~kHz, where $E_r = \hbar^2 k^2 / 2M$ is the photon recoil energy.
The term $U = N u$ represents the collective interaction, and normalization imposes $\sum_j |\psi_j|^2 = 1$.
Implicit in this applied form is the projection of the full interaction Hamiltonian onto the modes comprising the 1D synthetic lattice. This maintains mode-conserving (forward) collisions but ignores mode-changing collisions that would populate states outside of the considered synthetic lattice state space~\cite{An-Inter}. We note that the forward collision rate is dominant when $\rho \sigma v \ll g \rho$ (with $\sigma$ the scattering cross section and $v$ the relative velocity of scattering atoms), or equivalently when the relative $k$-vector of scattering atoms $\Delta k$ is much less than $\hbar/a$, which is well satisfied for most of the cases explored.

While these simple discrete-mode simulations adequately describe the results of Fig.~\ref{FIG:fig2}, they fail to capture several key features of the phase-driven dynamics data of Fig.~\ref{FIG:fig4}. Specifically, as discussed in the main text, discrete-mode simulations that ignore the real-space trapping potential both predict an effective reversal of atomic current (sign flip of the CPR, $\bar{n}$~vs.~$\phi$) under interaction strengths close to the self-trapping regime and also predict that the distribution width $\sigma_n$ should have $\phi$-dependence that is symmetric with respect to $\phi = 0$.

Figure~\ref{FIG:figS1} summarizes the predicted behavior under the discrete-mode simulations. Figure~\pref{FIG:figS1}{a,\,b} show the predicted curves for $\bar{n}$~vs.~$\phi$ and $\sigma_n$~vs.~$\phi$ for the simple discrete-mode NLSE, which assumes a uniform interaction energy of $U=h \times 1500$~Hz. The curves are evaluated at the same values of $J$ as considered in Fig.~\ref{FIG:fig4}, with the curve assignments specified by the color legend in panel (b). For comparison, we also include the results for the non-interacting dynamics (dashed line, for $U=0$).

While the non-interacting CPR in Fig.~\pref{FIG:figS1}{a} is purely sinusoidal, in line with our expectations of the group velocity of a simple cosine band structure, the interacting simulation curves become more and more skewed as $J$ is decreased. Furthermore, for the two smallest values of $J$ considered, the CPR in fact reverses sign over a broad range of $\phi$ values. Or, more directly, the CPR gains a more intricate $\phi$-dependence with several zero-crossings between $\phi = 0$ and $\phi = 2\pi$. To capture the dependence on the interaction energy-to-tunneling energy $U/J$, we summarize in Fig.~\pref{FIG:figS1}{c} the dependence of $\bar{n}$ on $\phi$ for the exemplary superposition state with relative phase $\phi = \pi/2$. For this discrete-mode GP simulation, a reversal of $\bar{n}$ (and by extension a reversal of the group velocity) is predicted for increasing $U/J$, along with an overall damping of the magnitude of $\bar{n}$ as full self-trapping sets in. This intriguing behavior is not observed in experiment, and is found to be absent in the simulations that incorporate the influence of the real-space trapping potential (detailed below).

For the distribution width dependence on $\phi$ shown in Fig.~\pref{FIG:figS1}{b}, one again finds the expected dependence in the non-interacting limit. That is, viewing this as probing the evolution under the cosine band structure, equivalent large widths are found when the dispersion is largest at the band center ($\phi = 0$) and band edge ($\phi = \pm \pi$), and the smallest widths are found at the points where the group velocity is nearly constant ($\phi = \pm \pi/2$). Interactions again significantly change the response of $\sigma_n$~vs.~$\phi$. In particular, the discrete-mode simulations clearly depict the $\phi$-dependent collapse of spreading that is seen in Fig.~\ref{FIG:fig4}. First, for decreasing $J$ (increasing $U/J$), one finds that the in-phase state with $\phi = 0$ undergoes a collapse of $\sigma_n$ while the spreading of the out-of-phase state ($\phi = \pm \pi$) remains robust. The interaction-dependence is summarized in Fig.~\pref{FIG:figS1}{d}, where we plot the dependence of $\sigma_n$ on the ratio $U/J$ for the superposition states with $\phi=0$ and $\phi = \pi$. Although these results are for a short evolution of just two tunneling times, there is an apparent difference in the responses of these two states, indicating a $\phi$-dependent self-trapping transition that occurs at lower $U/J$ values for the equal-phase superposition state as compared to the out-of-phase state.

Finally, we remark that further intricacies in the $\phi$-dependence of $\sigma_n$ can be seen in Fig.~\pref{FIG:figS1}{b} for the $J/h = 157$~Hz curve. Additional local minima of $\sigma_n$ are found (\textit{i.e.}, near $\pm\pi/2$), seemingly coinciding with the appearance of additional features in the $J/h = 157$~Hz CPR curve in Fig.~\pref{FIG:figS1}{a}. While the investigation of this behavior lies outside of the scope of the present study, this hints at the instability of different superposition states to a collapse of spreading, perhaps with an interpretation in terms of a nonlinear band structure and its evolution under an increasing $U/J$ ratio.

\subsubsection{Real-space Gross-Pitaevskii simulations}

As stated in the main text, a stiffer optical trapping condition was utilized for the data of Fig.~\ref{FIG:fig4} as compared to Fig.~\ref{FIG:fig2}, to achieve a slight enhancement of the collective $U$ (through enhancement of the atomic density). The higher trapping frequency of Fig.~\ref{FIG:fig4}, and the correspondingly shorter trap period, had resulted in the trapping potential having a non-negligible influence on the momentum-space evolution during the experiments. To capture the influence of the trap, along with the resulting inhomogeneous atomic density, real-space simulations were performed.

In these numerical simulations, we treat the system as a 1D BEC in a harmonic trap under proper controlling laser fields. The simulation is done with dimensionless units by choosing the recoil momentum $\hbar k=\hbar\times 5.91\times 10^{6}~\text{m}^{-1}$ and energy $E_r= h \times 2.03$~kHz as the units for momentum and energy, and setting $\hbar=1$. A complete run of simulations consists of the following three phases.

\begin{enumerate}
    \item
    \textbf{Imaginary-time evolution}. The BEC ground state is achieved through imaginary-time evolution with a time-split-operator numerical method~\cite{imag-time-split-op} and the ground-state energy is achieved to a precision up to $10^{-10}E_r$.
    \item
    \textbf{Equal superposition state}. The coupling laser field is incorporated in real space and the zero-momentum ground state is adiabatically evolved into the desired initial states with equally populated momentum modes. A relative phase between two major momentum states is imprinted through adjusting the phase of the lasers fields.
    \item
    \textbf{Wavepacket dynamics}. All real-space laser potentials are turned on, and the system is evolved (real-time evolution) up to two tunneling times before we achieve the final states.
\end{enumerate}

In our simulation, the real space is sampled by $2^{15}$ points and the step in momentum space is $6.67\times 10^{-3}\hbar k$. The time-evolution (both real and imaginary) step is $\delta t = 5\times 10^{-5}\hbar/E_r$ (corresponding to $3.92\times 10^{-9}$~s), which ensures the convergence of the final states. The trapping frequency is 60 Hz and the mean interaction strength of the initial state takes a value of $U/h=1568$~Hz as averaged over the whole trap. Finally, we confirm that our real-space GPE simulation agrees well with the tight-binding simulations when evaluated in the weak interaction and strong coupling limits.

For both the experimental data and simulation curves of Fig.~\ref{FIG:fig4}, the evaluations of the expectation value and standard deviation of the displaced position in the synthetic lattice are restricted to a range of states that does not span the entire 21-site lattice. Specifically, those evaluations are performed over the site ranges of $\{-6, \ldots 5\}$ for the $J/h = 1253$~Hz and 583~Hz sets, $\{-5, \ldots 5\}$ for the $J/h = 402$~Hz set, $\{-4, \ldots 4\}$ for the $J/h = 257$~Hz set, and $\{-3, \ldots 3\}$ for the $J/h = 157$~Hz set.

\end{document}